\documentclass[conference]{IEEEtran}
\IEEEoverridecommandlockouts
\usepackage{cite}
\usepackage{amsmath,amssymb,amsfonts}
\usepackage{algorithm2e}
\usepackage{graphicx}
\usepackage{textcomp}
\usepackage{xcolor}
\usepackage{float}
\usepackage{mathtools}

\usepackage{lipsum}
\usepackage{fancyhdr}

\def\BibTeX{{\rm B\kern-.05em{\sc i\kern-.025em b}\kern-.08em
    T\kern-.1667em\lower.7ex\hbox{E}\kern-.125emX}}

\newcommand\Tstrut{\rule{0pt}{3.2ex}}       
\newcommand\Bstrut{\rule[-0.9ex]{0pt}{0pt}} 
\newcommand{\TBstrut}{\Tstrut\Bstrut} 

\begin{document}

\title{\texttt{MARS}: Malleable Actor-Critic Reinforcement Learning Scheduler}

\author{\IEEEauthorblockN{ Betis Baheri\IEEEauthorrefmark{1}, Jacob Tronge\IEEEauthorrefmark{1}, Bo Fang\IEEEauthorrefmark{2}, Ang Li\IEEEauthorrefmark{2}, Vipin Chaudhary\IEEEauthorrefmark{3} and Qiang Guan\IEEEauthorrefmark{1}}
\IEEEauthorblockA{\IEEEauthorrefmark{1}Kent State University,
Kent, OH USA \\
\{bbaheri, jtronge, qguan\}@kent.edu
\\
\IEEEauthorrefmark{2}Pacific Northwest National Laboratory,
Richland, WA, USA \\
\{bo.fang, ang.li\}@pnnl.gov
\\
\IEEEauthorrefmark{3}Case Western Reserve University,
Cleveland, OH, USA \\
vxc204@case.edu
}
}

\IEEEoverridecommandlockouts
\IEEEpubid{\makebox[\columnwidth]{978-1-6654-4331-9/21/\$31.00 \copyright2022 IEEE \hfill}
\hspace{\columnsep}\makebox[\columnwidth]{ }}

\maketitle
\IEEEpubidadjcol

\begin{abstract}
In this paper, we introduce \texttt{MARS}, a new scheduling system for HPC-cloud infrastructures based on a cost-aware, flexible reinforcement learning approach, which serves as an intermediate layer for next generation HPC-cloud resource manager.
\texttt{MARS} ensembles the pre-trained models from heuristic workloads and decides on the most cost-effective strategy for optimization. A whole workflow application would be split into several optimizable dependent sub-tasks, then based on the pre-defined resource management plan, a reward will be generated after executing a scheduled task. Lastly, \texttt{MARS} updates the Deep Neural Network (DNN) model based on the reward. \texttt{MARS} is designed to optimize the existing models through reinforcement mechanisms. \texttt{MARS} adapts to the dynamics of workflow applications, selects the most cost-effective scheduling solution among pre-built scheduling strategies (backfilling, SJF, etc.) and self-learning deep neural network model at run-time.
We evaluate \texttt{MARS} with different real-world workflow traces. MARS can achieve  5\%-60\% increased performance compared to the state-of-the-art approaches. 

\end{abstract}


\hspace{1mm}
\section{Introduction}
As workflow applications grow in complexity,  Scientific Workflow Management Systems (SWMS's) have become essential components in recent HPC-cloud infrastructure
\cite{WorkflowManagementForWhom}.
Active research in scientific workflow management has enabled
systems used by scientists in practice, addressing many scientists' needs and improving system efficiency. 
Current workflow management systems, integrated with resource management systems, offer generic services to handle task management, distribution, monitoring and failure management on various types of platforms \cite{Distributed4683116,6655672}. 
Although workflow systems on cloud and HPC infrastructures have been studied with many services offering various capabilities, we still  lack optimized and sophisticated scheduler systems, which allow for collaboration of scientists running tasks on HPC systems and those running tasks on cloud systems.

Disparate workflows require different optimizations depending on the condition of the execution environment, state of hardware, resource management and task scheduling. For instance, CPU-intensive tasks need to be optimized to enhance the instruction throughput. 
Memory-intensive tasks should be scheduled in such a way as to minimize the use of global memory and only write back the final results; this configuration can be achieved by setting the proper configuration in the existing HPC resource manager.   

On the other hand, I/O intensive tasks should minimize the data transfers between different infrastructures.
In some cases, workflows require different resources. For example, part of the workflow can be executed on CPU-based HPC, and the rest would be benefit from GPU-based HPC or cloud infrastructure. In this case, the cost of executing the tasks of such workflow on two or more different cloud systems should be considered. 

Even though these challenges can be solved partially through meticulously designed heuristics, two or more of these factors should be considered for complex workflows. 
Pursuing recent research in HPC scheduling algorithms, the most common designs either apply an optimal solution for heuristic models or require changes at the system level 
that may need to replace the existing resource manager in the HPC system. This process must be repeated if the system workload changes or the metric of interest changes (e.g., more memory-intensive tasks than CPU-intensive tasks). 
The more appropriate solution is to use the existing resource manager by introducing an intermediate layer to create scheduling tasks.  
Following this architecture design, we can achieve a better optimal result without changing the system level resource manager. 

In summary, we illustrate the major challenges within  existing scheduling systems:
\begin{itemize}
    \item 
    The same scheduling strategy may not necessarily work for different infrastructures.
    For instance, in cluster scheduling, the execution time of a task varies with data locality, hardware health characteristics, interactions with other tasks, and interference on shared resources such as CPU caches, network bandwidth, etc.
    \item HPC system resources are usually managed by a resource manager,  e.g., SLURM. However, these tools are not optimized for dynamic changes in workflow performance characteristics. Accordingly, there are no optimizations for cost-effectiveness based on performance prediction. 
    \item 
    Practical instances have to make online decisions with noisy inputs and work well under diverse conditions. The decision between CPU, Memory, I/O and cost can have different meanings for individual workflow. Sub-optimization for an individual task may improve after running a couple of batches on the HPC system. 
    \item Scheduling policy switching between different systems can be challenging, workflow requirements and system configuration can be different on an individual HPC system. In most cases switching the scheduling system means re-initiating the HPC system, which requires time and termination of other users' tasks.   
    \item On basic workflows, optimization can be done using one dimension (CPU, Memory or I/O) optimization; there is no need for a multi-layer machine learning scheduler. The practical applications do not need further optimization. Simply using the existing resource manager would satisfy the users' needs.
    \item Lastly, workflows might benefit from various multi-level optimization and using machine learning scheduling techniques. In this case, each task has its own characteristics, which may be so sophisticated that we need to consider multi-dimension of the metrics of interest, such as IO, CPU and Memory.
\end{itemize}

To overcome these issues, we  design a generic scheduling system, enabling self-learning, performance adaptation and 
naturally working with the existing resource manager on HPC systems. 

In this paper, we introduce a malleable actor-critic reinforcement learning scheduler (MARS) to address the challenges within existing scheduling systems with the following features:
\begin{itemize}

    \item \texttt{MARS} presents a malleable scheduling policy ensembling A3C reinforcement learning and heuristic policies. 
    \item \texttt{MARS} optimizes scheduling performance through task parallelism and workflow classification through graph comparison and outperforms the state-of-the-art HPC schedulers  by 5-60\%. 
    \item  \texttt{MARS} requires none to minimal changes to the existing HPC resource manager such as Slurm and other cloud resource managers. From a design perspective, \texttt{MARS} is a good candidate for HPC-cloud heterogeneous environment.
    \item \texttt{MARS} design supports both simple and complex workflows, and the scheduling profile can be expanded from one to more dimensionality to bring more optimization on the desired characteristic.
    \item \texttt{MARS} requires none to minimal changes in users' workflows configuration, task optimization between user configuration and HPC system is done in the intermediate layer. Scheduling in such a way that we can benefit from simple optimization and complex machine learning schedulers. 
\end{itemize}

The rest of the paper is organized as follows: in section \ref{background} we discuss HPC workflow requirements and descriptions along with server parameters and our motivation. 
We explain how \texttt{MARS} integrates previous heuristic algorithms along with asynchronous actor-critic reinforcement learning, and we give a detailed explanation of the reinforcement learning approach and our decision on how to select the best suitable scheduling algorithm in section \ref{design}. 
We discuss our implementation methods in section \ref{implementation}. 
In section \ref{evaluation}, we discuss our results and compare them to previous works and present our observations. We also explain why \texttt{MARS} is outperforming the other approaches. Lastly, in section \ref{related_works} we discuss prior work in this area and conclude in  section  \ref{conclusion}. 

\section{Background AND Motivation}
\label{Back_Mot}
\subsection{Background}
\label{background}
A workflow application is a set of tasks or instructions executed on arbitrary input by particular order as steps. 
Workflow can be chained computation in physics, chemistry, etc. 
To improve the performance of workflows and create more meaningful relations between tasks, steps and requirements, we can use a directed acyclic graph (DAG) based on each component. As Hongzi M. and et al showed in their approach, creating DAG from workflows can be done in two categories. First, it can be done considering pure output related and their dependency. Second, another DAG can be generated based on tasks' resource requirements. 

Any workflow can have just one or multiple requirement DAGs based on the complexity of the workflow. Similarly, both DAGs can represent the target system resources and scheduling requirements. System resources are queries from existing resource manager which are explained in detail in section \ref{model}, scheduling requirements are the number of CPUs per node and/or entire workflow, the amount of memory, I/O and the cost based on desired parameters such as I/O throughput, CPU usage, GPU usage, etc. More details are provided in section \ref{design}. 

In order to better understand how to solve the complexity of both complicated and straightforward workflows scheduling, traditional optimization and machine learning techniques need to be studied. 
Mu'alem introduced optimization over First Come First Serve (FCFS) scheduling method, and AAhuva W. Mu’alem et al. \cite{932708}, their Backfilling method over well known FCFS algorithm was to overcome the fragmentation problem. Backfilling uses dynamic partitioning to schedule tasks on distributed systems to maximize performance. There are two major implementations, conservative Backfilling and Easy Backfilling. However, even though both methods were introduced to reduce starvation in the case of large workflows, both versions can cause starvation. In practice, if the workflow contains many tasks, it might be more beneficial to use machine learning techniques. In contrast, the Backfilling scheduling method would achieve better optimization when the workflows do not contain enough tasks to train and test the machine learning model.   

The volume of workflows tasks recently caused researchers to focus on machine learning techniques instead of traditional methods. One of the most recent and popular methods is reinforcement learning. The reinforcement learning (RL) method uses vector-based image transition. In order to translate the HPC scheduling parameters into this method, each resource, CPU, Memory, and I/O would be shaped into images. One requirement of this method is that the workflows` sizes must be the same. Then the translated images would be used to train the RL model. The optimal solution can be achieved after training the system. 

Knowing that both methods have limitations, we need to balance Backfilling and RL methods to support workflows more diversely.  The backfilling method suffers from optimization for large workflows, and the reinforcement technique needs initialization to be optimal, depending on workflow requirement and the volume scheduling algorithm needs to switch between these methods.

\subsection{Motivation}

The regular Reinforcement Learning (RL) schedulers require replacing existing HPC resource management tools, and in most cases, users have to adapt and change their workflow to satisfy the new system's requirements.
In a more specific explanation, HPC's existing resource manager would be replaced with an RL scheduler, and user configuration would have to change to use the new scheduler system. 

One reinforcement learning limitation is that the entire training set must be specific for the HPC system. Otherwise, the training model would not be optimized. Tuning hyperparameters and optimizing in favour of all dimensions is one of the limitations of RL methods.

Existing HPC resource managers suffer from large numbers of tasks. For better optimization, we need to either replace them with more specific algorithms depending on workflow or use an intermediate layer 
to communicate with the resource manager. Replacing the resource manager is time-consuming and requires knowledge of workflows. 
Since replacing the resource manager is costly and compromises support for legacy workflows, in our approach, we do not require to change the subsystem, and we use the existing tools to increase the performance\cite{HPC_RM_Book}. 

In our approach, we introduce a median layer to the existing HPC resource manager to avoid replacing the entire system and not depending on one solution for all possible cases. The user can specify how many parameter servers and nodes to use, including the amount of required resource (e.g. CPU, Memory, GPU, I/O, etc.), then submit the workflow to \texttt{MARS}. Our Scheduler \texttt{MARS} chooses between simple Backfilling or an advanced reinforcement learning (A3C) algorithm and then assigns tasks to nodes for execution by communicating with the existing resource manager on the HPC system.

\section{ \texttt{MARS} Design and Implementation}
\label{design}
\subsection{\texttt{MARS} System Overview}
\label{system_overview}
\begin{figure}[t]
    \includegraphics[width=1\linewidth]{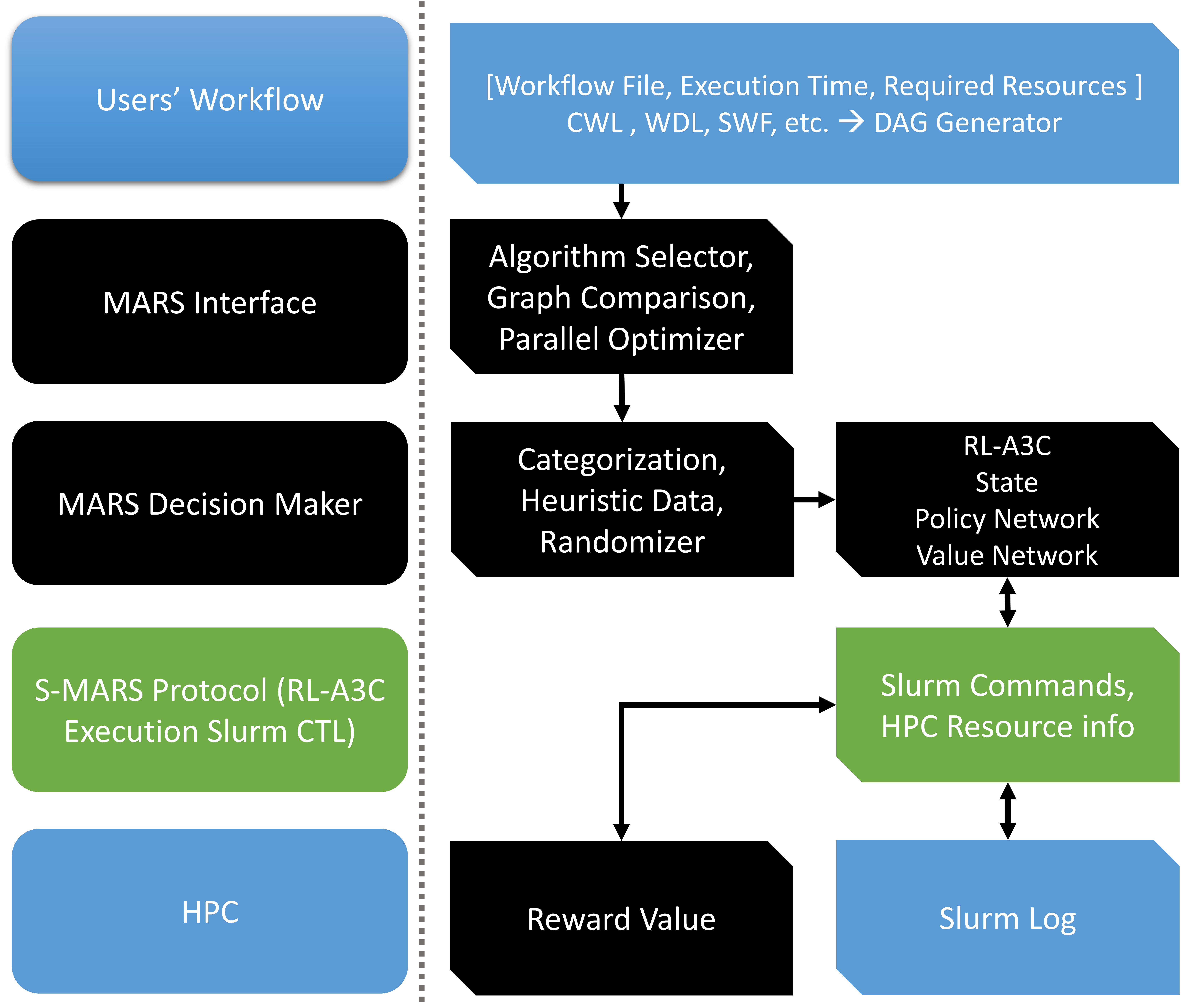}
    \caption{MARS System Overview}
    \label{fig:overview}
\end{figure}

Figure \ref{fig:overview} shows the overall system structure of \texttt{MARS}. We assume that the workflow description and generated DAG graph are provided to the scheduling system. One example of previous work done in \texttt{BEEFlow} \cite{841636600}, which proposed an in-situ analysis-enabled workflow management system that supports multiple platforms using HPC containers. 

Our design consists of several parts as shown in the figure \ref{fig:overview}, \texttt{MARS} Interface is the API that provides an interface to HPC users to submit their workflows on an HPC system. \texttt{MARS} decision-maker is in charge of deciding between Backfilling and RL-A3C algorithm, where algorithm selection and different optimizer can be assigned based on workflows requirement DAG. As shown, we use heuristic data, users' configuration, along randomization for unknown workflows characteristics. As mentioned before, large workflows can be optimized based on more than one dimension. In order to achieve the optimal solution, the RL-A3C needs to train on data, and in our case, a randomizer helps speed up this process and tune the hyper-parameters faster. More sophisticated methods such as population-based (PBT) in ML can surely help but, we observe it is unnecessary to use more complicated methods.
The rest of our design follows the RL-A3C principle with reward value read from the HPC resource manager, two policies, the model parameters from the DNN network, and the initial state of HPC resources. The Slurm commands and HPC resource information can be derived from existing resource managers such as Slurm, \texttt{MARS} uses the same commands to schedule tasks on an HPC system. 

In general, we take the following steps shown in algorithm \ref{alg_overview} to accomplish the optimization for each workflow.

\begin{algorithm}[t]

\caption{\texttt{MARS} Overall Algorithm}\label{alg_overview}
\SetAlgoLined
\KwResult{Saved Model M}
Input: Created DAG from workflows $\zeta$

Input : Decision D

Input: Policy $\nu$

Input: Available HPC Resources from Existing Resource Manager (SLURM) $HPC_R$

\eIf {Task $\iota$ and $\iota_{i+1}$ $!=$ dependency }
    {Compare and Parallel Tasks $\iota + \iota_{i+1}$}
    {
    
    D = $MARS_Decision$($\zeta$)
    
    $\nu$ = $MARS_Policy$(D)
    
    M = MARS($\zeta$,$\nu$,$HPC_R$)
    
    return M 
    
    }

\end{algorithm}
\setlength{\textfloatsep}{0pt}

The corresponding benefits to our design are:
\begin{itemize}
    \item Each workflow can be executed independently from others 
    \item HPC systems do not depend on a single algorithm 
    \item Workflows can run simultaneously with other workflows. In respect to users' workflows are not restricted or limited by the algorithms used for scheduling 
    \item Since the full optimization is done regardless of existing HPC systems. We can update saved models based on best suitable parameters
\end{itemize}

\subsubsection{\textbf{Algorithm Selection}}
\label{model}
In typical cases, resource management in the HPC system is based on CPU, memory, and I/O utilization. On the other point of interest, considering the cost of each task execution on other cloud infrastructures can help scientists minimize the overall cost.

Traditionally schedulers optimize tasks only on one dimension. A simple Backfilling scheduler can be an example. In Backfilling scheduling, the scheduler tries to optimize CPU usage.
In the next step, more sophisticated schedulers use modern Machine Learning algorithms to optimize tasks based on CPU, Memory and I/O.
However, in most methods, schedulers either sacrifice one feature for another or find the average solution. Recent ML schedulers use one specific reward function to update the trained model and learn from the previous execution.

In our proposal, \texttt{MARS} can adapt on different reward values read from the HPC resource manager and decide between a simple algorithm such as Backfilling to a more complicated online algorithm such as asynchronous actor-critic reinforcement learning to execute tasks. By creating a model based on the RL-A3C algorithm and updating that model with the similar technique that D. Zhang previously introduced and et al. \cite{2019arXiv191008925Z} we can reuse a trained model with similar workflows. However, the training of the system is highly correlated to the size and number of tasks in one arbitrary workflow. 

Based on our observation, small workflows such as a simple RNA search would be an inefficient model. On the other hand, complex and large workflows in RL, such as Blast, would cause an over-fitting of the network.
This phenomenon would result in an inefficient reward value and model. In our approach, by combining time window and custom loss function, the reward value and model generated from the workflow would be more accurate compared to previous approaches.

\subsection{Policy Model and Algorithm}
Our policy model depends on the size of the workflow. In terms of small workflows that can be optimized with the simple FCFS algorithm, \texttt{MARS} bypasses the RL algorithm and creates a simple schedule for tasks ready to be executed on HPC. On the other hand, when workflows contain a large subsection of tasks and the running time requires hours to days, \texttt{MARS} selects an arbitrary RL-A3C algorithm based on previously saved models.

The reinforcement learning module in \texttt{MARS} contains a scheduler agent, environment, and neural network based on server parameters input and reward value from the HPC environment. At each time step \textit{t} the agent observes the parameters on HPC state \textit{$s_t$}, then chooses an action \textit{$a_t$}. Following that action, the environment's state would proceed to \textit{$s_{t+1}$} and the agent receives reward \textit{$r_t$}. The state transitions and rewards are stochastic and are assumed to have the Markov property; i.e. the state transition probabilities and rewards depend only on the state of the environment \textit{$s_t$} and the action taken by the agent \textit{$a_t$}.

In most RL approaches, learning is done by performing \textit{gradient-decent} on the policy parameters. The critical idea in policy gradient methods is to estimate the gradient by observing the trajectories of executions obtained by following the policy. 
Similar to \textit{Monte Carlo Method} \cite{4736059}, samples are taken of multiple trajectories, and the empirically computed cumulative discounted reward is used. However, this approach is based on a naive algorithm and usually calculates a local maximum instead of the global maximum. In order to overcome this limitation, we use a similar method as other researchers 
, RL with Actor-Critic Algorithm (ACA) in \texttt{MARS}.

\subsubsection{\textbf{Reinforcement Learning Objects}}
\label{rl_objects}

Based on the definition for objective function for policy gradients, in our approach, parameters are read from the existing resource manager, and the action taken upon optimizing task execution is done by \texttt{MARS}.
Using well-known machine learning techniques \cite{2015arXiv150205477S,2017arXiv170706347S},  mapping between HPC server parameters and RL properties, we can redesign reinforcement learning to support HPC systems.
    
\subsubsection{\textbf{Reinforcement Learning Using Actor Critic }}
We can define the Actor-Critic method, where the Critic estimates the value function, which can be Q-Value or state value V-Value\cite{10.1137/S0363012901385691}. In our approach, we took the state value from an existing resource manager such as Slurm. \texttt{MARS} uses Slurm manager outputs to calculate the reward value. \ref{alg1} \texttt{MARS} Policy RL-A3C Algorithm:

\begin{algorithm}[h]
\SetAlgoLined
\KwResult{HPC Reward Estimation $\pi_\theta \approx \pi_*$}
Input: HPC Scheduling Action based on State Parameters $\pi(\textit{a}|\textit{s},\theta)$
Input: HPC CPU, Memory, I/O, Cost Values \v{v}$(\textit{s},\textit{w})$

Algorithm parameters: step sizes $\alpha^\theta$ \textgreater $0, \alpha^w$ \textgreater $0$

Initialize policy parameter $\theta \in \mathbb{R}^{d'}$ and state-value weights $w \in \mathbb{R}^d $(e.g., to 0) Set weights to 0 at beginning,

Initializing C as the Cost Probability added to evaluation;

 \While{for each epochs}{
  Initialize $S$ (first state of episode)\;
  $I \leftarrow 1$\;
  \While{$S$ is not terminal (for each time step)}{
   $A \approx \pi(\cdot | \textit{S}, \theta)$\;
   Take action $A$, Observe ${S'}$, $R$\;
   $\delta \leftarrow R + \gamma$ \v{v}$({S'},w) - $\v{v}$(S,w) $(if ${S'}$ is terminal, then \v{v}$({S'},w) = 0)$\;
   $w \leftarrow w + \alpha^w I \delta \nabla_w $\v{v}$(S,w)$\;
   $\theta \leftarrow \theta + \alpha^\theta I \delta \nabla_\theta ln \pi ( A | S, \theta)$\;
   $\theta \leftarrow \theta + \nabla C $\; 
   $I \leftarrow \gamma I $\;
   $S \leftarrow {S'} $\;
   
   }
 }
\caption{\texttt{MARS} RL-A3C Policy}\label{alg1}
\end{algorithm}
\setlength{\textfloatsep}{0pt}
As we explained earlier, the computation of the reward value can have different meanings.
The critic is a state-value function, \texttt{MARS} can be optimized based on Parameter Server values read from Slurm or any other resource manager, and final value results can be used to determine if there was an improvement or not.

\begin{figure*}[t]
\centering
    \includegraphics[width=0.9\textwidth]{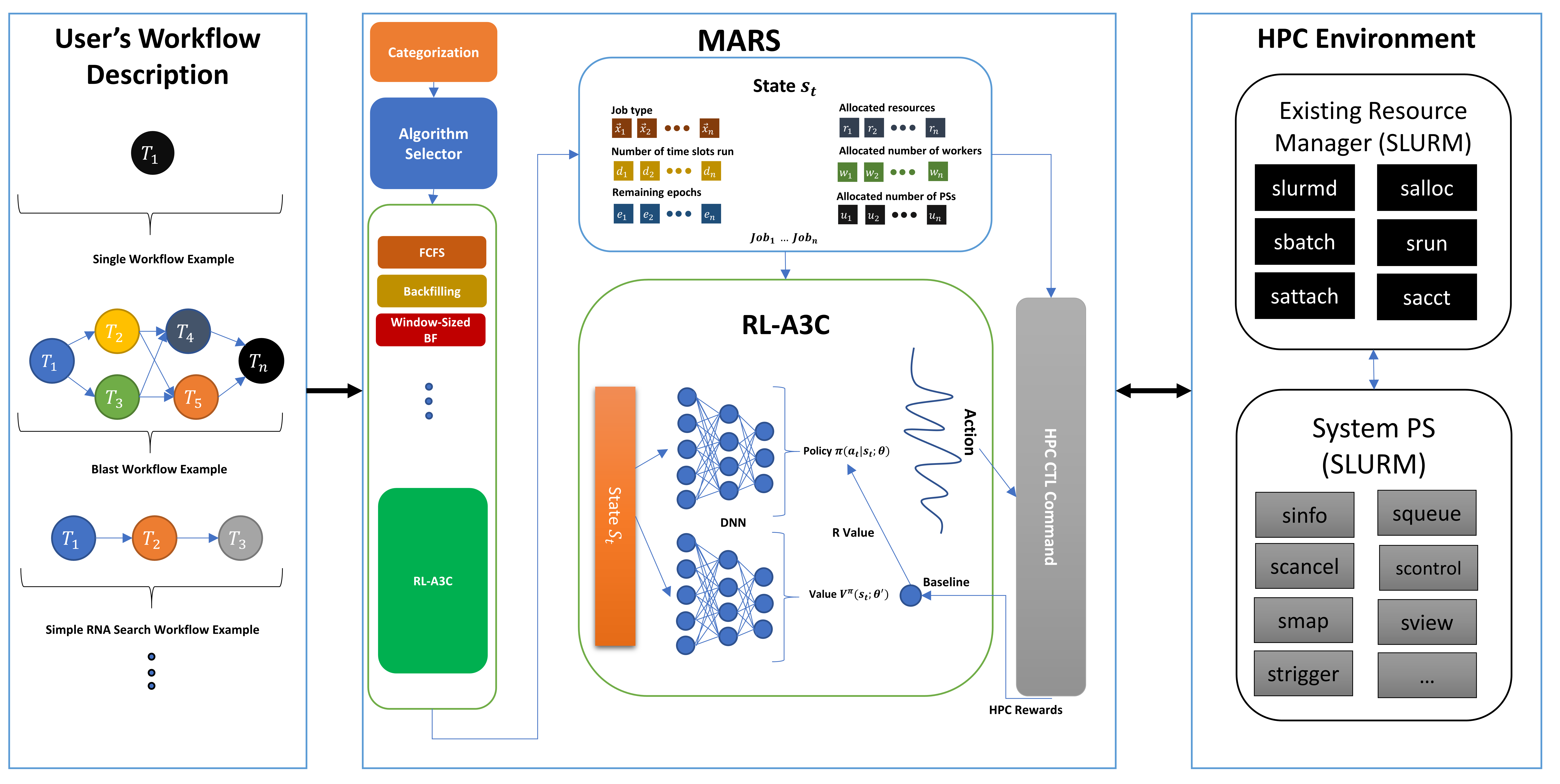}
    \caption{MARS Policy Network}
    \label{fig:policy}
\end{figure*}

Figure \ref{fig:policy} shows the Policy Structure of \texttt{MARS}. User's workflow description can be in any standard format such as Common Workflow Language (CWL), The Workflow Description Language (WDL), Standard Workload Format (SWF), etc. As mentioned before in section \ref{Back_Mot} preferably, the DAG is generated from workflow description containing tasks (tasks) to execute and the dependency between them. In our example, one workflow can be as simple as one task or have multiple dependent parts, such as the Blast example, or similar to a linear search workflow. The generated data then would be fed to our categorizing module, which determines the depth of the workflow based on the description, graph comparison algorithm and heuristic generated models. 

The algorithm selector module decides whether to use RL-A3C or basic FCFS, as mentioned before, for simple workflows which require only limited execution time. If no other workflows are running, and the description requires most system resources, running RL-A3C would cause overhead. However, in case \texttt{MARS} can combine multiple independent workflows and run RL-A3C, it would switch back to using the RL-A3C algorithm and build the best suitable model for that specific type. We kept the traditional algorithms such as FCFS, Backfilling, etc. 

In order to support legacy workflows and save on training time and in case an HPC system is not equipped with a GPU, a small optimization based on the known graph combining algorithm \cite{8723466} would run next to combine the parallel tasks. Compared to the standard Reinforcement Learning technique, we use this graph search algorithm to identify the best possible model to gain an optimal outcome and user input as a variable to differentiate between CPU, Memory, I/O, and Cost of each task. The generated model will train the system for optimization and feedback output. 

Next, \texttt{MARS} queries the available resources from Slurm, knowing the current state of the system and workflow description. Next \texttt{MARS} creates a state description based on Job type, the number of time slots run, remaining epochs, allocated resources on HPC, the number of workers based on the workflow description, and the number of parameters. Based on the previous discussion, we build a policy and value network, calculate a baseline, and initiate action; then, using the Slurm interface on HPC, we initiate a batch of tasks on HPC (Action). 

In addition, \texttt{MARS} needs to decide the best split between tasks and parallelism based on available resources, knowing that each workflow can be divided into sub-workflows based on searching paths, \texttt{MARS} categorizes tasks into groups. After this separation, it generates a deep neural network based on user input and CPU, Memory and I/O values. 

Finally, using Slurm CTL \texttt{MARS} queries about remaining available resources, current executing tasks, previously executed times, and corrupted previous tasks. \texttt{MARS} then calculates the reward value and uses a baseline. It updates the neural network. In order to overcome training overhead and inefficient models, \texttt{MARS} creates an arbitrary base network based on heuristic workflow data. If the data is absent from the database, we generate a similar workflow with smaller tasks to train the network.     

\subsubsection{\textbf{Graph Comparison and Parallel Optimizer}}
\label{graph_comparison}
In most RL-based schedulers, the generated workflow graph and cost are not considered. The deep neural network is purely based on workflow input data or previous execution. However, if we consider the graph generated from the workflow and use search algorithms to find the similarities in individual tasks, we can predict and categorize each task based on their CPU, Memory, I/O intensity. In addition, we can also consider the cost of each execution. Based on a predefined table, we can calculate how much each task would cost to run on some arbitrary cloud infrastructure.

In practice, Directed Acyclic Graphs (DAGs) have tens or hundreds of stages with different requirements and execution times. Based on the dependencies and requirements, each task can be executed in parallel or wait for other tasks to be completed. This complexity can be challenging in terms of scheduling, and to solve this issue \texttt{MARS} needs to execute tasks in parallel as much as possible without wasting CPU, or Memory utilization \cite{Graph_Example}.


As mentioned before, graph comparison is algorithmically hard, similar to C. Delimitrou and et al.  \cite{298205} approach, we use a scale-up and scale-out method to achieve the categorization. Assuming that the individual parts of a workflow's DAG can be categorized and compared to each other based on size and resources, \texttt{MARS} tries to combine the independent tasks as a single parallel task.  

\subsubsection{\textbf{Decision Making}}
\texttt{MARS} decision making is based on comparing the DAG and heuristic data, using a heuristic data model, DAG classification, or based on the size of the workflow \texttt{MARS} chooses the best suitable algorithm between basic back-filling and RL-A3C to execute an arbitrary workflow \ref{fig:policy}.
In complementing combining CPU, Memory, I/O and creating a general neural network, we generate an individual network based on graph comparison and user input for RL-A3C candidate workflows. Complementary to the previous method, the users' variable is used to determine the intensity of requirements and also, in order to achieve a better result, the logs from the target HPC system will be used in the evaluation. 

In the case of RL-A3C workflows, the first initiation and task execution would have to be on a more general deep neural network with a more straightforward reward function due to the lack of training data. However, after a couple of workflow executions, the first network can be replaced with a more complex network. After that process, \texttt{MARS} would get the output from the HPC system and calculate the universal reward means. As we know, returning a positive value from the reward function can identify the desired settings then and would cause \texttt{MARS} to continue optimizing on the same network for similar workflows. On the other hand, the cumulative negative reward value would cause a feature selection change in the network and update the loss function.

Algorithm \ref{alg_system} shows the basic decision making of the \texttt{MARS} scheduler. Our design uses workflow size and configuration to decide on the algorithm policy.  In our experiment, we observe that workflows with a size less than 512 are not sufficient to run directly on RL-A3C. In order to improve this issue, we either combine the following workflow with the previous one or run the heuristic algorithm. In the algorithm's first part, we combine the following workflow with the current workflow. Next, if the compatibility of dimension fails or the existence of the following workflow is absent, then \texttt{MARS} chooses the heuristic algorithm. Next, for the large workflows, we split those into sub-workflows and execute the RL-A3C algorithm to avoid over-fitting the network. In each step, we save the RL-A3C model for future use.  

\begin{algorithm}[h]
\SetAlgoLined
\KwResult{Best Suitable Action $\alpha$}
Input: Workflow $ \chi$ \& Workflow size $\eta$

Initializing workflow task size, Queue, Task, Model: $ \eta \leftarrow \chi $ , Q, $\omega$, M 

\eIf{$\eta$ \textless MEDIAN} {

\eIf{$ \chi_{i+1} == TRUE$ \& $ \chi_{i+1}$ is compatible (RL-A3C vector dimensions) with $\chi_i $ }{

$\chi = \chi_{i} + \chi_{i+1}$ \textgreater MEDIAN ;

$Q \leftarrow \eta$;

$M \leftarrow MARS-RL-A3C(Q)$;

}
{

\eIf{$\eta$ \textless MIN }
    {
    $Q \leftarrow \eta$;
    
    SJF(Q);
    
    }
    {
    $Q \leftarrow \eta$;

    UNICEF(Q); 
    
    }
 }   
    $M \leftarrow MARS-RL-A3C(Q)$;
    
}{

\While{$\eta$ \textgreater MAX}{

$\omega = \frac{\omega}{2}$

$Q \leftarrow \omega$;

$MARS-RL-A3C(\omega)$ 

$M \leftarrow MARS-RL-A3C(Q)$;

}

$Q \leftarrow \omega$;

$MARS-RL-A3C(\omega)$ 

$M \leftarrow MARS-RL-A3C(Q)$;
}
\caption{\texttt{MARS} Decision Making Policy}\label{alg_system}
\end{algorithm}
\setlength{\textfloatsep}{0pt}




\section{Implementation}
\label{implementation}

The \texttt{MARS} algorithms are implemented using Tensorflow \cite{45381} and Gym OpenAI \cite{gym}. For the training process we used Proximal Policy Optimization (PPO) algorithm derived from OpenAI Spinning Up library \cite{SpinningUp2018,2017arXiv170706347S}.

We used a randomly generated data set based on real workflows and actual real-world data from different sources to evaluate the proposed solution. The real-world workflows are based on SWF archive data as shown in Table \ref{tab:workloads}. 
\begin{table}[H]
\caption{List of Workload Traces}
\label{tab:workloads}
\centering
\begin{tabular}{llll}
 \hline
Name                                        & \ CPU & Month(s) & Date \\ \hline
SDSC IBM-SP2  & 128       & 24       & 1998          \\ 
SDSC IBM-Blue & 1152      & 32       & 2000          \\ 
High Performance Computing Center     & 240       & 42       & 2002          \\ 
Argonne National Laboratory Intrepid        & 163840    & 8        & 2009          \\ 
Synthetic\_G001                             & 256       & 12       & 2019          \\ 
Synthetic\_G002                             & 1024      & 6        & 2019          \\  \hline

\end{tabular}

\end{table}

In our experiment, we aim to compare the previous works with \texttt{MARS}. 
We compare \texttt{MARS} with heuristic job scheduling algorithms, shown in Table \ref{tab:algorithms}. The table \ref{tab:algorithms} shows the heuristic scheduling policies infused with \texttt{MARS}, which can improve the performance of legacy and modern workflows. \texttt{MARS} is compared with two well-known policies: First Come First Served (FCFS), where the arrival order schedules tasks; and Shortest Job First (SJF), where tasks
with shorter processing times are scheduled ahead of the other tasks. Some other comparative policies are  WFP3, and UNICEF \cite{5289206}, which are based on the processing time, requested number of
cores and waiting time of the tasks. WFP3 favours shorter and older tasks over large ones without starvation, and UNI favours small tasks by using a fast turnaround policy for performance enhancement. Policy F1, F2, F3, and F4 \cite{10.1145/3126908.3126955} represent the nonlinear machine learning-based scheduling algorithms for minimizing the average bounded slowdown of tasks. Based on our observation, switching to known heuristic algorithms and RL-A3C increases the performance and saves a noticeable amount of time in training for the basic legacy workflows.

\begin{table}[h]
\caption{Heuristic Scheduling Policy Used}
\label{tab:algorithms}
\centering
\begin{tabular}{ll}\hline
Name   & Function                                            \\\hline
FCFS   & \(ABS(t)= s_t \)                                      \\
SJF    & \(ABS(t)= r_t \)                                       \\
WFP3   & \(ABS(t)= - (w_t / r_t)^3 *n_t\)    \\
UNICEP & \(ABS(t)= - w_t/(log_2(n_t) *r_t)\)                 \\
F1     & \(ABS(t) = log_{10}(r_t) * n_t + 8.70 * 10^2 log_{10} (s_t) \) \\
F2       & \(ABS(t) = \sqrt{r_t} *n_t + 2.56 * 10^4 *log_{10}(s_t) \)                                                    \\
F3       &  \(ABS(t) = r_t * n_t + 6.86 * 10^6 log_{10} (s_t) \)                                                   \\
F4       &  \(ABS(t) = r_t * \sqrt{n_t} + 5.30 * 10^5 log_{10} (s_t) \)                                                  
\end{tabular}

\end{table}

In an HPC  system, workflow tasks may arrive continuously. In order to train the model using RL-A3C, we save the training results after a predefined window time, and then we let the actor-critic algorithm improve the model. 
After building a basic model based on the RL algorithm, the Actor-Critic part evaluates the network. This strategy would create a training batch for the workflow. If the batch size is too small, \texttt{MARS}' decision module gives two options if the remaining workflow size is sufficient enough \texttt{MARS} combine sub-workflows. On the other hand, in the absence of sufficient size, \texttt{MARS} would switch back to back-filling or FCFS algorithm.

In our experiment running basic workflows on RL-A3C takes a significant amount of time to train and causes inefficiency in HPC systems. In order to overcome this issue, a combination of legacy and RL-A3C algorithms would be more appropriate. Another issue in RL-A3C is over-fitting the model due to the large batch size and exponential growth of the number of possible tasks. In order to solve this issue, we introduce a median layer to create sub-workflows. Based on our observation, the best training sets are between 512 to 20000 running on 2000 to 4000 epochs for RL-A3C. Knowing that the smaller or larger batch sizes could introduce an issue, the \texttt{MARS} decision module would combine or split the sub-workflows. 

As we described in Section \ref{design}, in RL-A3C, the state is the input of the DNN agent, and the representation of state is a vector containing available resources and pending tasks.    
In HPC number of pending and arriving tasks can vary. However, in DNN, the vector to create the network should be fixed-sized. In order to overcome this issue, we took the same approach as previous works and added extra 0s to the end of the vector \cite{2018arXiv181001963M}.

\section{Evaluation}
\label{evaluation}
\begin{figure*}[ht]
    \includegraphics[width=\textwidth]{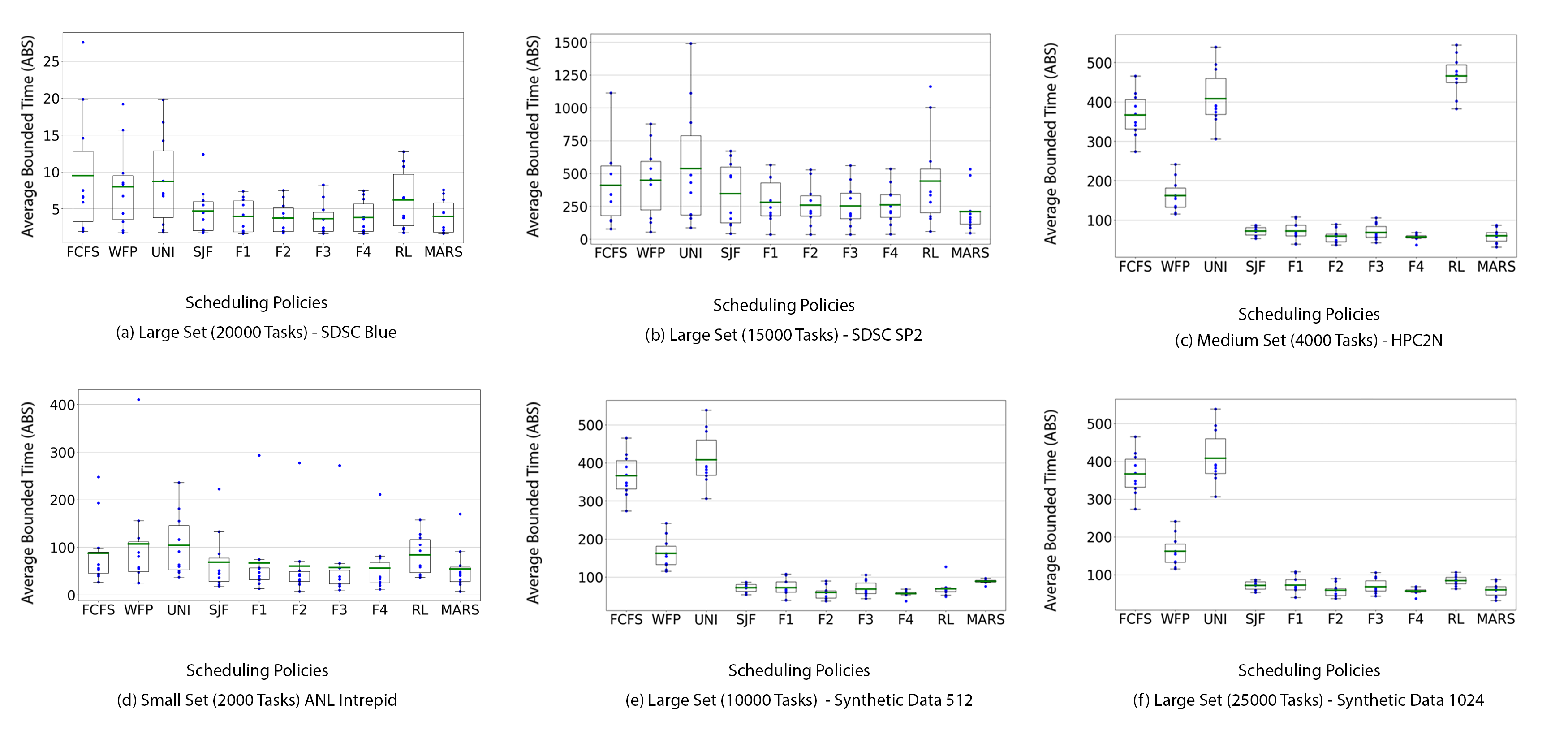}
    \caption{Performance of scheduling policies with different workload traces.}
    \label{fig:algorithm_comp}
\end{figure*}

In this section, we present our results obtained by running \texttt{MARS} scheduler on a simulated environment using data traces generated from HPC data centers. First, we describe the environment setup and workflow traces used in our experiment, then we evaluate different algorithms and compare them to our approach. We discuss the performance evaluation under different conditions and workloads of HPC environments. Our simulator was inspired by a similar method used by D. Zhang et al. \cite{2019arXiv191008925Z}. However,
to comply with our approach, we extended the simulator with Gym and OpenAI to return the proper reward values from the environment. Running the training set on an actual HPC environment requires an enormous number of iterations to learn, considering that most HPC environments are not capable of running the RL-A3C algorithm due to lacking GPU capability or available resources for non-HPC applications. The best approach is to either dedicate an arbitrary external server to train the model or run the simulation in a local environment. 

\subsection{Simulation Environment}
We simulate a homogeneous HPC environment executing tasks based on moving forward the timestamp instead of running those tasks. These workflows were based on traces collected from real systems, but we use the CWL and SWF workflows formats to guarantee compatibility. When a workflow is generated, if the resources required to run an arbitrary task belonging to the generated workflow are not present, the simulator uses the back-filling method to run smaller tasks first. 

\begin{figure*}[ht]
    \includegraphics[width=0.9\textwidth]{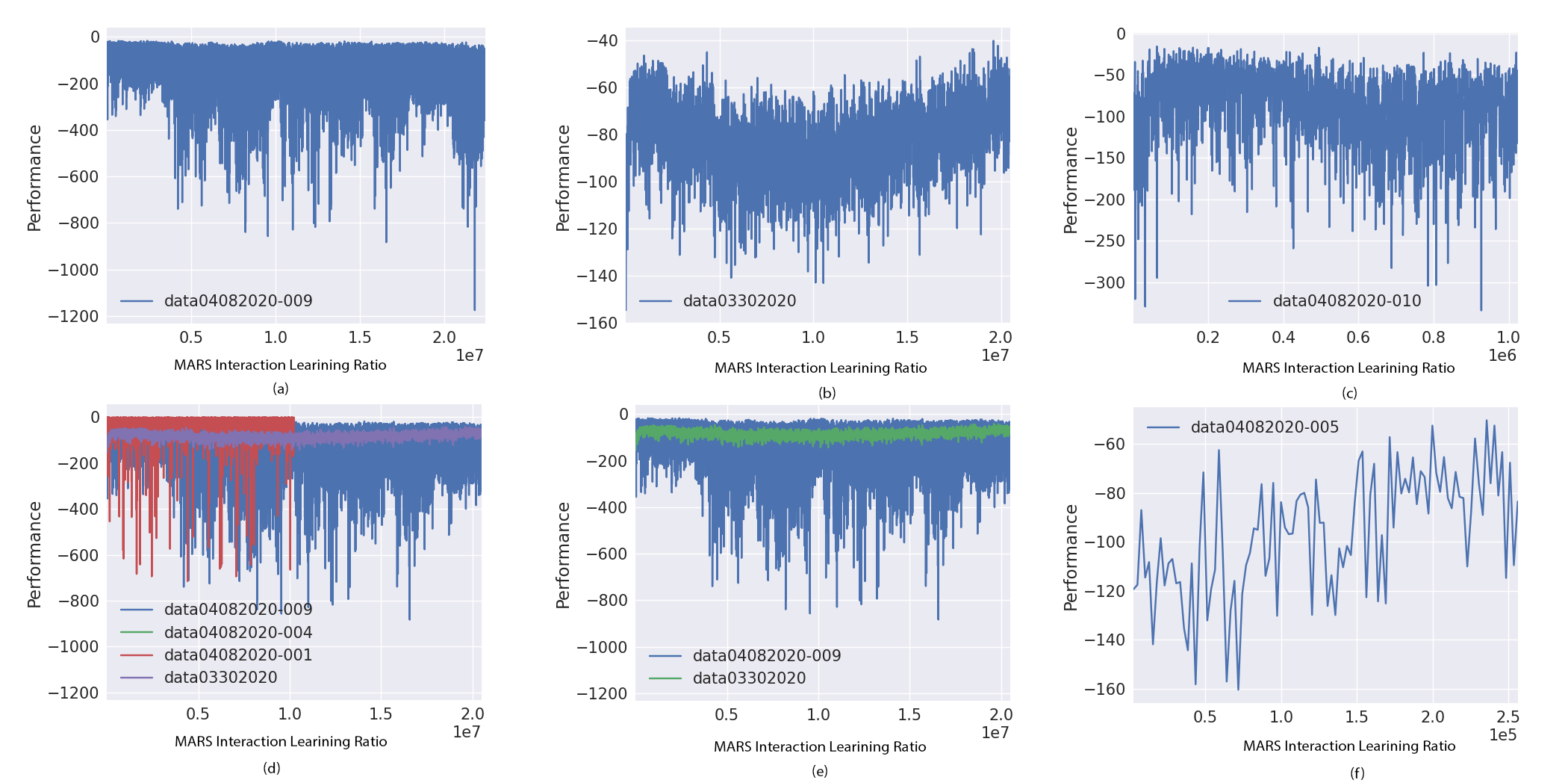}
    \caption{Performance of MARS under different learning ratios}
    \label{fig:learning_comp}
    \vspace{-4mm}
\end{figure*}

\subsection{HPC Reward and Metrics}

HPC scheduling metrics are mostly based on response time, and it is defined as the total wall-clock time from the instant at which the task is submitted to the system until it finishes its run. The most basic method to calculate the running time and wait time for tasks is slowdown, \( slowdown = \frac{T_w + T_r}{T_r}\). A more sophisticated method is to take the average slowdown to minimize the wait time\cite{10.5555/646382.689681}. 
Table \ref{tab:metrics} shows different evaluation metrics. The problem with the slowdown metric is that it overemphasizes the importance of short jobs; to overcome this issue, Feitelson et al. \cite{bounded-slowdown} have suggested Bounded-slowdown. The behaviour of this metric depends on the
choice of \(\tau\), which is the threshold value.  Zotkin and et al. \cite{10.5555/822084.823251} have introduced a new problem where tasks that do the same amount of work with the same response time may lead to different slowdowns results due to their shape, which is the ratio of processors to time. This introduces another metric known as a per-processor slowdown. We used average bounded slowdown instead of per-processor because, in our workflow examples, the shape of our test systems are identical to each other.  
\begin{table}[h]
\caption{Scheduling Metrics $T_r$ is the execution time of the job, $T_w$ is the time spent in turnaround \cite{805303}}
\label{tab:metrics}
\centering
\begin{tabular}{|c  c|}\hline
Metric   & Formula                                            \\
 \hline 
Slowdown   & \( \frac{T_w + T_r}{ T_r} \)                    \TBstrut                  \\
Bounded-slowdown    & \( max\{ \frac{T_w + T_r}{ max\{ T_r,\tau \} },1\} \)            \TBstrut                           \\
pp-slowdown   & \( max\{ \frac{T_w + T_r}{ P * max\{ T_r,\tau \} },1\}\)    \TBstrut \\
\hline
\end{tabular}
\end{table}
In our approach we set the goal as minimizing the average bounded slowdown \(  = - max \{ \frac{T_\omega + T_r}{max \{ T_r,\tau \} } , 1 \} (-ABS) \). At the start of the algorithm, calculating the average is not possible, instead we return 0 as a reward. After finishing the entire task sequence then the RL-A3C agent gets the average as \( -ABS \). 
\subsection{Results}
In this section, we show that \texttt{MARS}, by using a combination of heuristic and the RL-A3C algorithms, can improve the performance, time and avoid over-fitting the network for scheduling tasks on HPC systems.
Most reinforcement learning algorithms need to be configured with proper parameters from HPC. Figure \ref{fig:algorithm_comp} shows the different policies based on different configurations, where the y-axis is the average bounded slowdown, and the x-axis is the different scheduling policies. 

Our scheduler ratio of training and testing was 70\% to 30\%, similar to most other RL algorithms. We categorized three different configurations and sizes for our testbed: the small data-set contained between 512 to 2000 tasks, the medium size data-set was from 2000 to 9000 tasks, and lastly, the large data-set was between 10000 to 25000 tasks. We randomly selected tasks from different data sets and performed experiments with different configurations. We considered the number of iterations per task in DNN and the delay between task arrival. By experimenting with different configurations, we showed that the proper configuration causes a significant difference in result in reinforcement learning and heuristic algorithms. Lastly, we added the cost-aware probabilities after creating the RL-A3C model.

In figure \ref{fig:algorithm_comp} part (a), we choose a large data-set from IBM SDSC Blue with 20000 tasks to train and 6000 tasks to test. However, since the data configuration was chosen randomly, the reinforcement learning algorithm reacts worst than \texttt{MARS}. Similarly, in part (b), we selected 15000 random tasks and observed the same result; however, if the workflow size is large enough and the data is consistent with the configuration of DNN, the RL-A3C algorithm will improve. Figure \ref{fig:algorithm_comp} part (c) was HPC2N data-set with 4000 selected tasks, and Figure \ref{fig:algorithm_comp} part (d) contains small selected tasks from ANL Intrepid data-set. All three experiment configurations were chosen randomly. 

As discussed, the \texttt{MARS} scheduler tries to solve this issue in two ways: it either combines the tasks to generate a proper size for training and testing in RL-A3C or switches back to a heuristic algorithm. In our experiment, we showed that in the case of a proper and ideal configuration \ref{fig:algorithm_comp} (e), RL-A3C performs better compared to \texttt{MARS}. However, since in HPC, achieving the ideal configuration is rather difficult, in other cases, such as Figure \ref{fig:algorithm_comp} part (f), using the suggested method derives a better performance. Our experiment shows \texttt{MARS} on average can achieve between 5\% to 60\% better performance compared to other policies.  

Another issue in reinforcement learning to consider is over-fitting the network. In figure \ref{fig:learning_comp} we observe that based on data-set configuration and learning ratio, we can achieve different performances. Figure \ref{fig:learning_comp} part (a) is a large data-set with 50000 iterations per task, which causes RL-A3C learning to interact frequently with the HPC system.   

Figure \ref{fig:learning_comp} part (b) is the optimal configuration with the proper size data-set; however, in part (c), the configuration and HPC parameters change randomly, and that causes the RL-A3C agent to interact with HPC more often. Figure \ref{fig:learning_comp} part (d) and part (e) shows the comparison of different experiments together, and lastly, part (f) shows an insufficient data-set size to train. To resolve these issues, \texttt{MARS} tries to update the reward values from HPC after each iteration, and by selecting a heuristic algorithm for small data-set sizes, we bypass the inefficient training model.

In our test experiment, the cost of each task was randomly generated, and after RL-A3C soft-max values, we incorporate costs as another probability function as a probability between 0 and 1. We used Gaussian distribution to add the cost factor to the final step of the DNN soft-max calculation. As discussed before, adding the cost to the training model would result in a unique data model. As a consequence of keeping the model's generality, the cost would be incorporated after creating the DNN network. A more specific reward value can be derived from the HPC system by calculating the cost with each action taken by the agent. As shown in figure \ref{fig:algorithm_comp}, with random configuration for RL-A3C, the performance decreases between 5\% to 60\%. However, by using \texttt{MARS} policy and combining heuristic and RL-A3C with cost-awareness, the performance improves back to an optimal solution. 

\section{Related Work}
\label{related_works}
HPC task scheduling has been a long-time research topic. Countless studies have been done, including heuristic algorithms such as First Come First Serve (FCFS), Shortest Job First (SJF) and more sophisticated policies like WEP3, UNICEF and even machine learning approaches. \texttt{MARS} is clearly different
from the existing studies as it takes advantage of existing resource management on HPC systems and it combines the best suitable algorithm to maximize the performance and reduce the training time\cite{5289206,10.1145/3126908.3126955}. 

Mirhoseini et al. \cite{46646,2017arXiv170604972M} use DRL to optimize placement of computation graph, Xu et al. \cite{2018arXiv180105757X} use the same method to select routing paths between network nodes for traffic, and Mao et al. \cite{10.1145/3098822.3098843} used the same principle to select video stream rates dynamically. 

Recently, several studies also started to leverage deep reinforcement learning in resource allocation and job scheduling in a distributed environment, such as DeepRM\cite{10.1145/3005745.3005750}, and Decima\cite{2018arXiv181001963M}. However, none of these uses existing HPC resource management and combines the heuristic algorithm with deep reinforcement learning. 

Although they used similar DRL methods as \texttt{MARS}, these studies are not designed for scheduling HPC tasks, which are fixed, rigid, and non-preemptable.

These differences led to different designs and optimizations in \texttt{MARS}, detailed in Section \ref{system_overview}. The most recent HPC tasks scheduling \cite{10.1145/3126908.3126955} uses brute force simulations to generate a large number of data samples, each of which shows the best scheduling decision given a random job sequence. Then, applying machine learning methods on these data samples to build scheduling functions that can best fit these samples.

\section{Conclusion}
\label{conclusion}
In this study, we proposed a new cost-aware reinforcement learning policy for task scheduling on HPC systems using the existing resource manager, allowing the system administrators and users to optimize the scheduling of tasks based on any preferred algorithm and cost-effectiveness. We showed that using \texttt{MARS}, which combines heuristic and deep reinforcement learning actor-critic algorithm, HPC systems can be optimized for both legacy and complex workflows. We performed better  by choosing different configurations and switching between heuristic and RL-A3C. 
\texttt{MARS} can improve the modularity and support for both legacy and complex workflows, and it can optimize task execution based on the most appropriate approach.

\bibliographystyle{IEEEtran}
\bibliography{bibliography} 

\end{document}